\documentclass[reprint,amsmath,amssymb,aps,pra]{revtex4-1}
\usepackage{graphicx}% Include figure files
\usepackage{dcolumn}% Align table columns on decimal point
\usepackage{bm}% bold math
\usepackage{hyperref}% add hypertext capabilities
\usepackage{amsmath} % split in equation
\usepackage{color}

\begin{document}

\title{Noise-resistant phase gates with amplitude modulation}

\author{Yan Wang$^1$}
\author{Jin-Lei Wu$^1$}
\author{Jin-Xuan Han$^1$}
\author{Yong-Yuan Jiang$^{1,2,3,4}$}
\author{Yan Xia$^5$}
\author{Jie Song$^{1,2,3,4}$} \email{E-mail: jsong@hit.edu.cn}
\affiliation{$^1$ School of Physics, Harbin Institute of Technology, Harbin, 150001, China\\
	$^2$ Key Laboratory of Micro-Nano Optoelectronic Information System, Ministry of Industry and Information Technology, Harbin, 150001, China\\
	$^3$ Key Laboratory of Micro-Optics and Photonic Technology of Heilongjiang Province, Harbin Institute of Technology, Harbin, 150001, China\\
	$^4$ Collaborative Innovation Center of Extreme Optics, Shanxi University, Taiyuan, 030006, China\\
	$^5$ Department of Physics, Fuzhou University, Fuzhou, 350002, China}

\date{\today}

\begin{abstract}
We propose a simple scheme for implementing fast arbitrary phase gates and employ pulse modulation to improve the gate robustness against different sources of noise. Parametric driving of a cavity is introduced to induce Rabi interactions between the cavity and qutrits, and then a two-qubit arbitrary phase gate is constructed by designing proper logical states. On this basis, we implement amplitude-shaped gates to obtain enhanced resilience to control errors in gate time and frequency detuning. By virtue of specifically designed logical states, the scheme displays intrinsic resistance to the energy relaxations of qutrits. Furthermore, we show that the gate robustness against cavity decay can be enhanced significantly with amplitude modulation.

\end{abstract}
\maketitle
\section{Introduction}
Constructing a quantum computer requires sequences of accurate gate operations on a large number of quantum bits (qubits) \cite{figgatt2019parallel}. Any unitary transformation, including multiqubit gates, can be decomposed into a series of elementary two-qubit gates together with single-qubit operations in principle \cite{PhysRevA.52.3457}. Recently, much attention has been attracted on realizing universal two-qubit controlled-not gates and controlled-phase gates in various physical systems including trapped ions \cite{leibfried2003experimental,PhysRevLett.122.030501,PhysRevA.95.022328,PhysRevA.74.032322,PhysRevA.65.064303,PhysRevLett.90.217901}, neutral atoms \cite{PhysRevLett.83.5166,PhysRevX.8.011018,PhysRevLett.104.010503,PhysRevApplied.13.024059,Shen:19,Wu:20},  linear optics \cite{PhysRevLett.95.210505,PhysRevLett.106.013602}, and superconducting circuit quantum electrodynamics (QED) \cite{PhysRevLett.116.180501,PhysRevLett.124.120501,PhysRevA.100.062324,PhysRevA.96.052317,PhysRevA.90.012328}. In addition to these atom-based (both natural and artificial atoms) gates, photon-based gates enabled by atom-mediated interactions have also been extensively studied. Recent examples include the designs of single–photon Hadamard gates \cite{10.1117/12.2517090} and two–photon controlled–phase gates \cite{10.1117/12.2517035,PhysRevA.90.012328} on the solid-state platforms. Among various physical architectures, circuit QED has recently become a leading platform for quantum computing as well as quantum information processing (QIP). One of the outstanding features of superconducting circuit is that the coupling strength between a superconducting qubit and a microwave field can be artificially engineered \cite{GU20171}. On this basis, tunable qubit-resonator coupling has been realized experimentally using a tunable mutual inductance between a phase qubit and a lumped-element resonator \cite{PhysRevLett.104.177004}, or using a three-island transmon qubit with a tunable dipole moment  \cite{PhysRevLett.106.083601,PhysRevLett.106.030502,PhysRevB.84.184515}. Based on the context of circuit QED, high two-qubit gate fidelity above $99\%$ has been demonstrated in recent experiments  \cite{barends2014superconducting,PhysRevA.93.060302}. 

Despite rapid progress in realizing high-fidelity gate over the years, gates fidelity remains a major bottleneck for scalable QIP as well as fault-tolerant quantum computation. The gate fidelity is generally limited by two sources of noise, namely control errors in gate parameters and decoherence induced by noisy environment. The former can arise due to  slow drifts in experimental parameters such
as the resonator frequency, or imperfections in the control apparatus as well as imperfect operations. The latter is caused by the inevitable interaction between the system and environment \cite{PhysRevA.98.053830,Chen:17}, which will collapse the quantum state and make the quantum information no longer correct, thereby compromising the power of quantum computation \cite{PhysRevLett.79.1953,PhysRevA.101.032329}.  Although fault-tolerant quantum error correcting is capable of counteracting a certain degree of errors or decoherence, it cannot be brought into play unless the underlying error rates are small to begin with \cite{PhysRevA.65.042303,PhysRevLett.98.190504}. Therefore, seeking for high-fidelity gate operations that are robust in the presence of noises or errors becomes an imperative. 
 
In this article, we present a simple and experimentally feasible method to implement fast arbitrary phase gates in a qutrit-cavity coupled system, introducing amplitude modulation to make the operation more resilient to the two sources of noise described above. Specifically, the features and advantages of the present work are summarized as follows. The scheme employs a parametric drive to squeeze the cavity mode, which induces Rabi interactions between the qutrits and cavity with exponentially improved effective coupling strength. Then two-qubit arbitrary phase gates can be implemented by designing proper logical states and parameter conditions. Since the gate time is inversely proportional to the effective qubit-cavity coupling, it can be shortened significantly by employing strong parametric drive. In contrast to previous gate schemes based on the dispersive regime where the desired effective interactions are derived using perturbation theory with large-detuning conditions  \cite{PhysRevLett.116.180501,Yang:18,PhysRevA.86.024301,Han:20,Han_2019}, this proposal enables faster implementation of the phase gates. Besides, we show that the gate robustness against parameter errors, including gate timing error and frequency detuning error, can be increased by shaping the qutrit-cavity coupling strength with time-dependent amplitude modulation. In principle, the error suppression is achieved by means of optimizing the phase-space trajectory of the cavity mode for minimizing residual qubit-cavity entanglement in case of errors. Furthermore, the proposed amplitude-shaped gates are robust to the energy relaxations of qutrits due to the coding of quantum imformation on the qutrits ground states and in particular, exhibit significantly enhanced robustness against cavity decay compared to the unshaped gate. 
%Our proposal is well-suited for contemporary circuit QED technology, and could.

The remainder of the paper is structured as follows. In Sec.\,\ref{sec2}, we describe the physical model and demonstrate in detail how to implement the two-qubit arbitrary phase gate. We then take the $\pi$-phase gate as an example for numerical simulations to verify the gate performance. In Sec.\,\ref{sec3}, we propose to use amplitude modulation to improve the gate robustness against parameter errors. Then, the effects of decoherence caused by the energy relaxations of qutrits and cavity decay on the shaped gates are discussed. Finally, we briefly discuss the possible experimental implementation and summarize our conclusions in Sec.\,\ref{sec4}.

\begin{figure}
	\includegraphics[width=1\columnwidth]{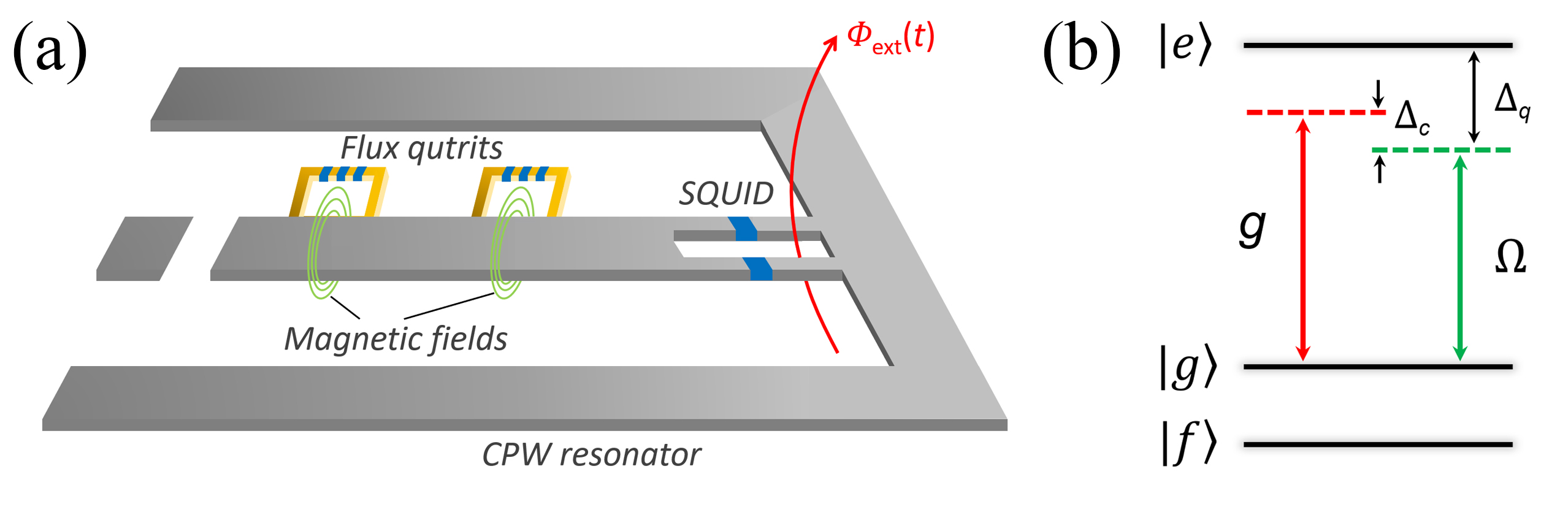}
	\caption{(a) Schematic diagram of a potential realization of the proposed scheme in circuit QED where two flux qutrits are coupled to a coplanar waveguide (CPW) resonator via the induced magnetic field. A flux-pumped superconducting quantum interference device (SQUID) connected to the resonator implements the parametric drive. (b) Energy level structure and transitions in a qutrit. 
	}
	\label{fig1}
\end{figure}

\section{Construction of two-qubit arbitrary phase gate}\label{sec2}
We consider a system containing two qutrits coupled to a single-mode cavity that is subjected to a parametric drive. Figure \ref{fig1}(a) illustrates a potential implementation of the system in the context of circuit QED \cite{PhysRevLett.103.147003,PhysRevLett.105.023601,RevModPhys.85.623}. The two qutrits have identical level structure, i.e., an excited state $\lvert e\rangle$ and two ground states $\lvert g\rangle$ and $\lvert f\rangle$ [see Fig. \ref{fig1}(b)]. The transitions between $\lvert e\rangle$ and $\lvert g\rangle$ are coupled to the cavity mode with coupling strength $g$. A classical field (with frequency $\omega$) is imposed on the qutrits to drive the transitions between $\lvert e\rangle$ and $\lvert g\rangle$ with Rabi frequency $\Omega$. The state $\lvert f\rangle$ is not affected during the interaction. Particularly, a parametric (two-photon) driving field of frequency $\omega_p$ and amplitude $\Omega_p$ is introduced to drive the cavity, which produces a nonlinear process equivalent to degenerate parametric amplification. For the current model we assume a sufficiently narrow bandwidth of the generated photon pair centered on the cavity frequency, which is much less than the scale of amplitude $\Omega_p$ to ensure that the model is valid.
As shown in Fig. \ref{fig1}(a), this parametric drive of the resonator can be implemented by modulating the flux $\Phi_{\mathrm{ext}}(t)$ threading the SQUID loop connected to the middle of the transmission
line \cite{macklin2015near,doi:10.1063/1.2964182,Zhong_2013}.

A direct effect of the parametric drive is that it will result in exponentially enhanced qutrit-cavity interactions that can be well-described by a usual Rabi Hamiltonian \cite{PhysRevLett.120.093602,PhysRevLett.120.093601,PhysRevA.100.012339,PhysRevA.99.023833}. To be more specific, the Hamiltonian of the system described above in a proper observation frame is given by ($\hbar=1$, hereinafter)
\begin{equation}\label{H0}
\begin{split}
H=&\Delta_c a^\dag a + \sum_{j=1,2}\Big\lbrack\Delta_q\lvert e\rangle_j\langle e\lvert+g(\lvert e\rangle_j\langle g\lvert a + a^\dag \lvert g\rangle_j\langle e\lvert)\\&
-\frac{\Omega}{2}(\lvert e\rangle_j\langle g\lvert+\lvert g\rangle_j\langle e\lvert)\Big\rbrack -\frac{\Omega_p}{2}(a^2+a^{\dag 2}).
\end{split} 
\end{equation}
Here, $a^\dag (a)$ is the creation (annihilation) operator of the cavity mode with frequency $\omega_a$; The detunings are $\Delta_c=\omega_a-\omega_p/2$, $\Delta_q =\omega_e-\omega_g-\omega_p/2$, where $\omega_z$ is the frequency associated with level $\lvert z\rangle$ ($z=e,g$); We have also assumed $\omega=\omega_p/2$. By diagonalizing the cavity-only part in $H$ with unitary $U_s=\exp[r_p(a^2-a^{\dag 2})/2]$, where the squeeze parameter $r_p$ is defined via $r_p=(1/2)\mathrm{arctanh}(\Omega_p/\Delta_c)$, the Hamiltonian (\ref{H0}) is rewritten as 
\begin{equation}\label{Hs}
\begin{split}
H_s=&\Delta a^\dag a + \sum_{j=1,2}\Big\lbrack \frac{g}{2}e^{r_p}(a + a^\dag)(\lvert e\rangle_j\langle g\lvert+\lvert g\rangle_j\langle e\lvert)\\&
-\frac{g}{2}e^{-r_p}(a^\dag-a)(\lvert e\rangle_j\langle g\lvert-\lvert g\rangle_j\langle e\lvert)\\&-\frac{\Omega}{2}(\lvert e\rangle_j\langle g\lvert+\lvert g\rangle_j\langle e\lvert)\Big\rbrack,
\end{split} 
\end{equation}
where $\Delta=\Delta_c\mathrm{sech}(2r_p)$ and we have set $\Delta_q=0$ for simplicity. Apparently, the first term in the summation has the form of Rabi Hamiltonian. The second term $H_{\mathrm{Err}} \equiv \sum_{j}\left[ -\frac{g}{2}e^{-r_p}(a^\dag-a)(\lvert e\rangle_j\langle g\lvert-\lvert g\rangle_j\langle e\lvert)\right] $, indicating the deviation from ideal Rabi Hamiltonian, could be neglected in a large amplification limit $e^{r_p}\to\infty$ (which requires $r_p$ to be sufficiently large, see Appendix for detailed discussion), such that $H_s$ after moving to the interaction picture with respect to $H_0=\Delta a^\dag a$ is transformed to
\begin{equation}\label{Hsprime}
\begin{split}
H_s'=g_s S_x(a e^{-\mathrm{i}\Delta t} + a^\dag e^{\mathrm{i}\Delta t})-\Omega S_x,
\end{split} 
\end{equation}
where $g_s=g e^{r_p}$ and $S_x=\sum_{j}(\lvert e\rangle_j\langle g\lvert+\lvert g\rangle_j\langle e\lvert)/2$. 

The coupled system described by $H_s'$ undergoes a unitary evolution with the propagator being  $U(t)=e^{-\mathrm{i}F(t)S_x a} e^{-\mathrm{i}G(t)S_x a^\dag} e^{-\mathrm{i}A(t)S_x^2} e^{-\mathrm{i}B(t)S_x}$, where we have
\begin{equation}\label{int}
\begin{split}
F(t)=&g_s\int_{0}^{t}  e^{-\mathrm{i}\Delta t'} dt'=\frac{\mathrm{i}g_s}{\Delta}(e^{-\mathrm{i}\Delta t}-1),\\
G(t)=&g_s\int_{0}^{t} e^{\mathrm{i}\Delta t'} dt'=\frac{\mathrm{i}g_s}{\Delta}(1-e^{\mathrm{i}\Delta t}),\\
A(t)=&\mathrm{i} g_s\int_{0}^{t} F(t') e^{\mathrm{i}\Delta t'} dt'=\frac{-g_s^2}{\Delta}(t-\frac{e^{\mathrm{i}\Delta t}}{\mathrm{i}\Delta}+\frac{1}{\mathrm{i\Delta}}),\\
B(t)=&-\int_{0}^{t} \Omega dt'=-\Omega t.
\end{split}
\end{equation}
As can be inferred from $U(t)$, a maximally two-qubit entangled state emerges in the $S_z$ basis for $\Delta t=2\pi$, $A(t)=\pi/2$ and $\Omega=0$, which corresponds to the M$\o$lmer-S$\o$rensen gate operation \cite{PhysRevA.62.022311,PhysRevLett.121.180501,PhysRevLett.121.180502}. In the $S_x$ basis, $U(t)$ could implement a two-qubit phase gate coded via proper logical states. To this end, we assume that the ground state $\lvert f\rangle$ of qutrits is coded to carry the logical state 0 and superposition state $\lvert +\rangle=(\lvert e\rangle + \lvert g\rangle)/\sqrt{2}$ to carry the logical state 1. By setting $\Delta \tau=2k_1\pi$ (where $\tau$ is the gate time and $k_1$ a positive integer), we have $F(\tau)=G(\tau)=0$. Then the propagator can be written as merely acting in the qubit subspace, which is reduced to $U(\tau)=e^{-\mathrm{i}A(\tau)S_x^2} e^{-\mathrm{i}B(\tau)S_x}$, with $A(\tau)$ and $B(\tau)$ being $A(\tau)=-2k_1\pi g_s^2/\Delta^2$ and $B(\tau)=-2k_1\pi \Omega/\Delta$, respectively. It can be easily found that
\begin{equation}\label{evolution}
\begin{split}
&\lvert f\rangle_1 \lvert f\rangle_2 \stackrel{U(\tau)}{\longrightarrow}\lvert f\rangle_1 \lvert f\rangle_2,\\
&\lvert f\rangle_1 \lvert +\rangle_2 \stackrel{U(\tau)}{\longrightarrow}e^{-\mathrm{i}\frac{A(\tau)}{4}}e^{-\mathrm{i}\frac{B(\tau)}{2}}\lvert f\rangle_1 \lvert +\rangle_2,\\
&\lvert +\rangle_1 \lvert f\rangle_2 \stackrel{U(\tau)}{\longrightarrow}e^{-\mathrm{i}\frac{A(\tau)}{4}}e^{-\mathrm{i}\frac{B(\tau)}{2}}\lvert +\rangle_1 \lvert f\rangle_2,\\
&\lvert +\rangle_1 \lvert +\rangle_2 \stackrel{U(\tau)}{\longrightarrow}e^{-\mathrm{i}A(\tau)}e^{-\mathrm{i}B(\tau)}\lvert +\rangle_1 \lvert +\rangle_2.
\end{split}
\end{equation}
Further, we demand that $A(\tau)/4+B(\tau)/2=-2k_2\pi$ and $A(\tau)+B(\tau)=-(2k_3\pi+\varphi)$ where $k_2$ and $k_3$ are integers and satisfy $k_3\ge2k_2-1/2 $ (see below). In this case, Eq. (\ref{evolution}) becomes
\begin{equation}\label{evolution1}
\begin{split}
&\lvert f\rangle_1 \lvert f\rangle_2 \stackrel{U(\tau)}{\longrightarrow}\lvert f\rangle_1 \lvert f\rangle_2,\\
&\lvert f\rangle_1 \lvert +\rangle_2 \stackrel{U(\tau)}{\longrightarrow}\lvert f\rangle_1 \lvert +\rangle_2,\\
&\lvert +\rangle_1 \lvert f\rangle_2 \stackrel{U(\tau)}{\longrightarrow}\lvert +\rangle_1 \lvert f\rangle_2,\\
&\lvert +\rangle_1 \lvert +\rangle_2 \stackrel{U(\tau)}{\longrightarrow}e^{\mathrm{i}\varphi}\lvert +\rangle_1 \lvert +\rangle_2,
\end{split}
\end{equation}
which conducts a two-qubit arbitrary phase gate. In the following, we take the gate with $\varphi=\pi$ as an example to discuss detailedly the gate performance. The corresponding constraints become $A(\tau)=-(2k_3+1-4k_2)\pi$ and $B(\tau)=(2k_3+1-8k_2)\pi$, which result in $\Delta=g_s/\sqrt{(2k_3+1-4k_2)/k_1}$ and $\Omega=-(2k_3+1-8k_2)\Delta/2k_1$, respectively. The gate time is thus inversely proportional to the effective coupling $g_s$, i.e., $\tau\sim 1/g_s$. Since $g_s$ increases significantly with enhancing the parametric drive, which could be orders of magnitude larger than the original coupling, the gate operation could be attained at a very short time.

\begin{figure}
	\includegraphics[width=0.9\columnwidth]{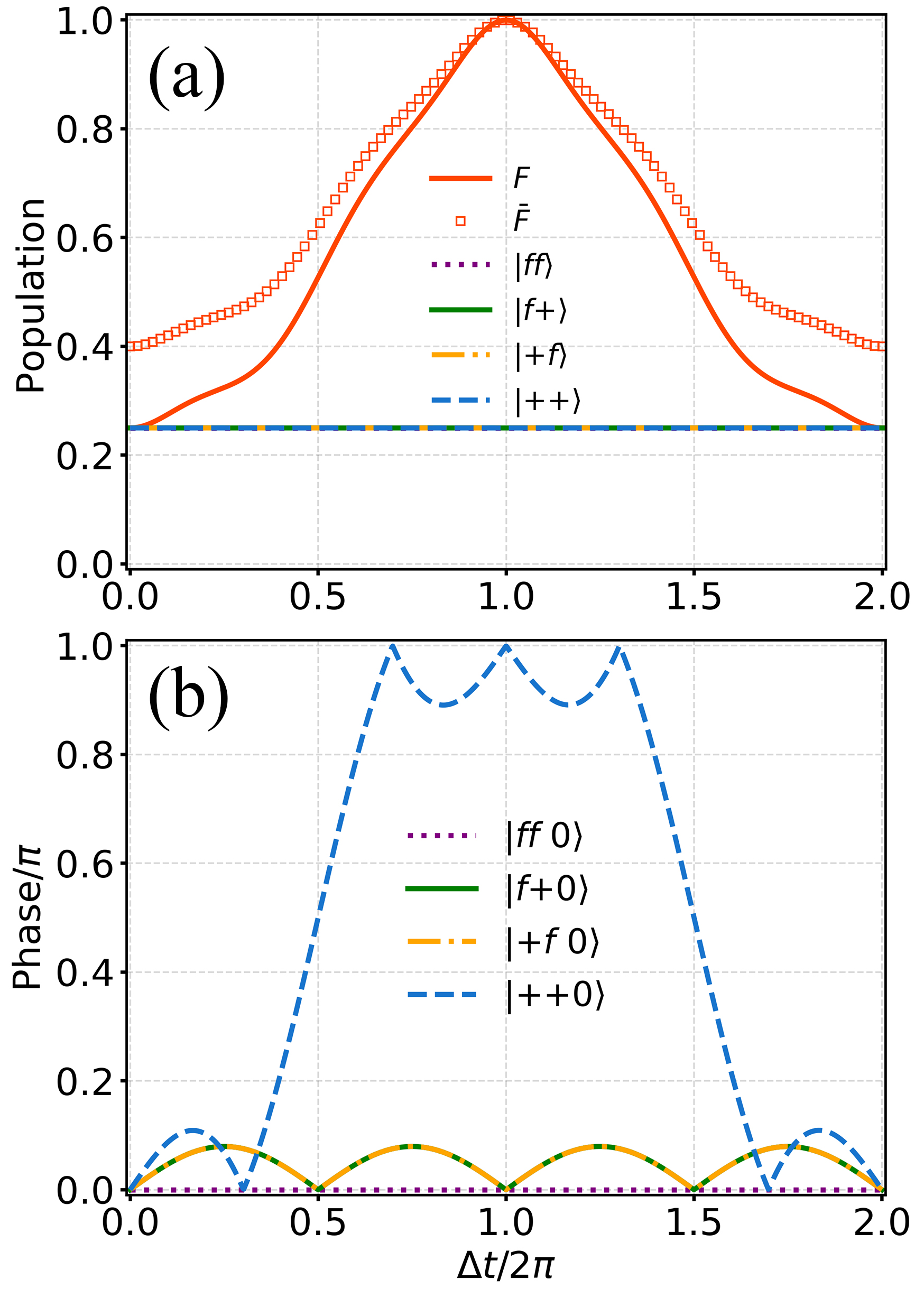}
	\caption{(a) Evolutions of the four computational states' population and initial-state-specified (average) gate fidelity $F$ ($\bar{F}$). (b) Phase evolutions of the four computational states. Parameters used here: $r_p=2.5$, $\Delta=g_s$ and $\Omega=-\Delta/2$.
	}
	\label{fig2}
\end{figure}

To show clearly the gate operation, we plot in Fig. \ref{fig2}(a) the evolutions of the four computational states' population and the gate fidelity \cite{JOHANSSON20121760,JOHANSSON20131234}, which are obtained by numerically solving the Liouville equation with respect to the full Hamiltonian given by Eq. (\ref{Hs}). Here we choose the initial state of the system to be $\lvert\psi_0\rangle=\frac{1}{2}(\lvert ff\rangle+\lvert f+\rangle+\lvert +f\rangle+\lvert ++\rangle)\otimes\lvert 0\rangle$ where the latter state refers to initial cavity vacuum. The gate fidelity is defined via $F=\langle \psi_{\tau}\lvert\rho_q(t)\lvert \psi_{\tau}\rangle$ in which $\lvert\psi_{\tau}\rangle=U(\tau)\lvert\psi_0\rangle$ and $\rho_q(t)$ is the reduced density operator of the system after tracing out the cavity degree of freedom. Also, we plot the phase evolutions of the four computational states in Fig. \ref{fig2}(b). It can be found in Fig. \ref{fig2} that the populations of the four computational states remain unchanged during the gate evolution, while $\lvert ++0\rangle$ obtains an accumulated phase of $\pi$ at the gate time $\tau=2\pi/\Delta$ ($k_1$=1 and $k_2=k_3=0$). Correspondingly, the gate fidelity reaches above 0.9999 at the instant $\tau$. 
For an experimentally feasible coupling $g/2\pi=50$ MHz \cite{niemczyk2010circuit}, the gate time is estimated approximately as 0.16 ns.

In the numerical simulations above, we assume a particular two-qubit superposition state to evaluate the gate fidelity. Strictly speaking, using only a specific initial state is insufficient to test the gate performance. In view of this, we introduce the average fidelity metric that takes into account all possible initial inputs, which is defined based on a trace-preserving quantum operator, as\cite{NIELSEN2002249}
\begin{equation}\label{ave fid}
\bar{F}(\varepsilon,U)=\frac{\sum_{l=1}^{4^{N+1}} \mathrm{tr} \left[ U u_{l}^{\dag} U^\dag\varepsilon(u_l)\right] +d^2}{d^2(d+1)},
\end{equation}
where $u_{l}$ is the tensor of Pauli matrices $II, I\sigma_x, \dots , \sigma_z \sigma_z$ in the computational basis $\{\vert f\rangle, \vert +\rangle\}$, $U$ is the perfect phase gate, $d = 2^{N+1}$ for a $(N+1)$-qubit gate, and $\varepsilon$ is the trace-preserving quantum operation obtained through
solving the master equation. Based on Eq. (\ref{ave fid}), we plot the average gate fidelity using the same set of parameters as above, as depicted by the hollow squares in Fig. \ref{fig2}(a). One can clearly see that the average fidelity at the gate time is also close to unity (precisely, above 0.9999). For simplicity, we use the initial-state-specified gate fidelity $F$ to characterize the gate against noises in the following sections.

\section{Robustness against control errors and decoherence}\label{sec3}
\subsection{Amplitude-shaped phase gates}
For implementing quantum gates in practice, perfect control of experimental operations is desired for achieving high fidelity, which requires accurate calibration and short-term stability of the gate parameters. This is obviously difficult in real experiments due to imprecise apparatus and imperfect operations. For our scenario, a conspicuous drop of the fidelity can be observed for slight drifts in gate time $\tau$, as depicted in Fig. \ref{fig2}(a). In addition to the gate time, inaccuracies and drifts of other gate parameters (e.g., the detuning $\Delta$) will also reduce the gate fidelity. Therefore, seeking for effective schemes to improve gate robustness against control errors is of growing significance.

Inspired by pulse shaping or modulation techniques employed extensively in recent studies \cite{nphys3967,Felix2018,PhysRevLett.123.260503,PhysRevLett.123.100501,PhysRevA.100.043413,Han:20,PhysRevA.91.032325,PhysRevA.97.033407}, we herein propose to use shaped time-dependent coupling  $g(t)$ to replace the constant $g$, which is designed as $g(t)=g_m \sin^2(\alpha t)$, with $g_m$ and $\alpha$ being maximum amplitude and suitable constants, respectively. Experimentally, amplitude modulation of the qubit-cavity coupling has been demonstrated in circuit QED system 
\cite{Vepsineneaau5999,PhysRevX.5.021027}. At the gate time $\tau$, a ``soft" end is desirable \cite{PhysRevLett.123.260503}, which signifies $\alpha \tau=m\pi$ for an integer $m$ identifying the number of pulses present in the envelope of $g(t)$. Without loss of generality, we pick up the case with $m=1$ in what follows, i.e., $\alpha \tau=\pi$. Then, by substituting $g(t)$ into Eq. (\ref{int}) and solving analytically the integrals, we obtain 
\begin{equation}\label{ints}
\begin{split}
F'(t)=&e^{r_p}\int_{0}^{t} g(t') e^{-\mathrm{i}\Delta t'} dt'=\frac{\mathrm{i} 2e^{r_p}g_m \alpha^2 (e^{\mathrm{-i}\Delta t}-1)}{\Delta(4\alpha^2-\Delta^2)},\\
G'(t)=&e^{r_p}\int_{0}^{t} g(t') e^{\mathrm{i}\Delta t'} dt'=\frac{\mathrm{i} 2e^{r_p}g_m \alpha^2 (1-e^{\mathrm{i}\Delta t})}{\Delta(4\alpha^2-\Delta^2)},\\
A'(t)=&\mathrm{i}e^{r_p}\int_{0}^{t} F'(t') g(t') e^{\mathrm{i}\Delta t'} dt'=\frac{e^{2r_p}g_m^2 \widetilde{A}(t) }{8(\Delta^3-4\Delta \alpha^2)^2},\\  
\end{split}
\end{equation}
with
\begin{equation}
\widetilde{A}(t)=3\Delta^5 t-20\Delta^3\alpha^2 t+\mathrm{i}32\alpha^4(e^{\mathrm{i}\Delta t}-\mathrm{i}\Delta t-1).
\end{equation}

\begin{figure}
	\includegraphics[width=0.9\columnwidth]{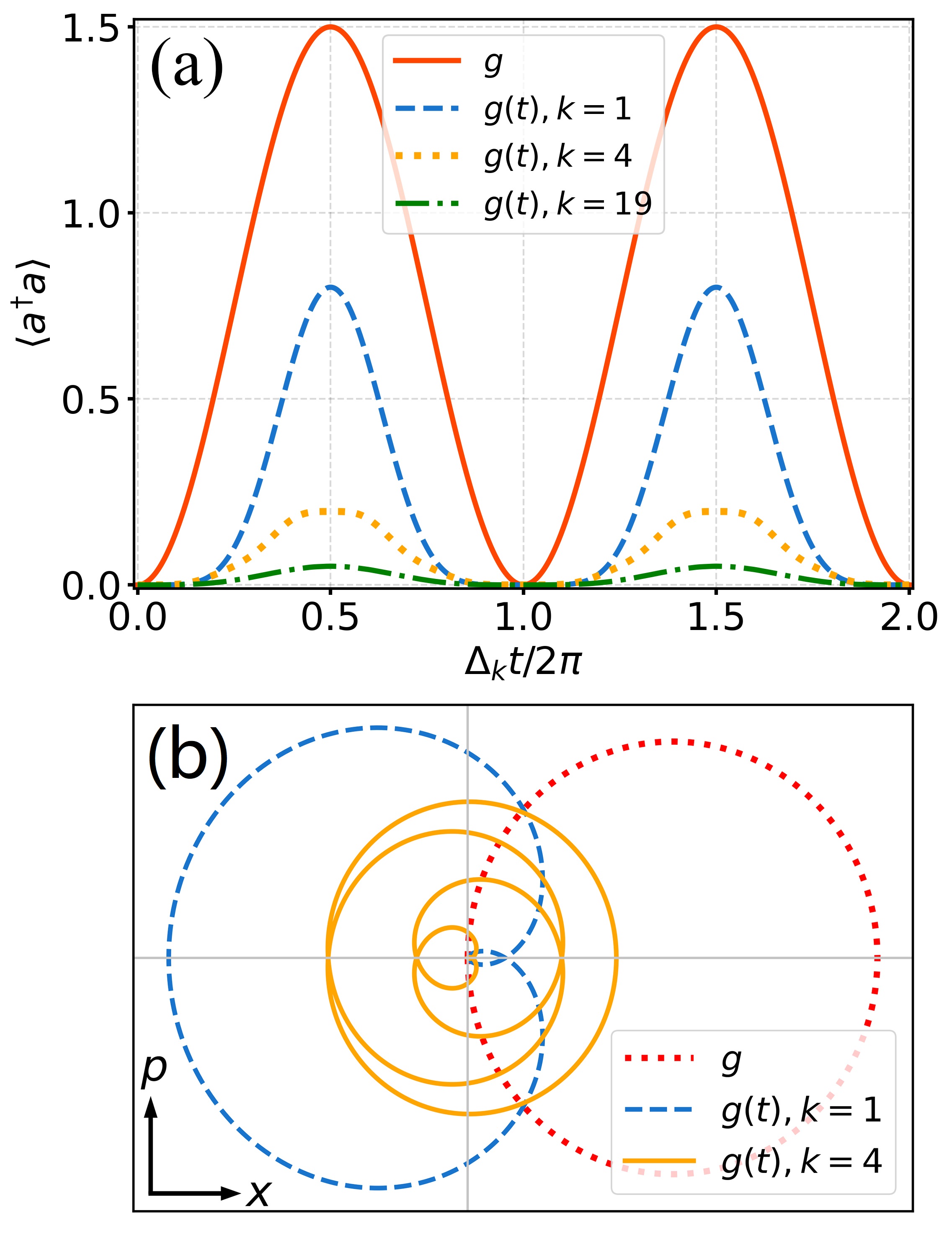}
	\caption{(a) Time evolutions of the mean photon number and (b) phase space trajectories during the unshaped gate (with constant $g$) and the shaped gates (with time-dependent $g(t)$) operations. The parameters used for the unshaped gate are the same as in Fig. \ref{fig2}. The parameters used for the shaped gates: $g_m=g$; $\Delta_k$, $\alpha_k$ and $\Omega_k$  satisfy the constraints given in the text; To make the shaped gate times close to that for the unshaped gates, $r_p$ for $k=1$, 4 and 19 are set as 3.28, 3.78 and 4.49, respectively. The common initial state of the system is $\lvert\psi_0\rangle$.
 	}
	\label{fig3}
\end{figure}

\noindent In Eq. (\ref{ints}) one finds that multiple sets of $\Delta$ and $\tau$ (or $\alpha$) can produce the required gate phase $A'(\tau)=-2\pi$, each of which satisfies $\Delta_k \tau_k=2(k+1)\pi$ where $k$ is a positive integer defined herein as the order of the shaped gate and $\tau_k$ the gate time for this order. In addition, the other constraint $B(\tau)=\pi$ requires $\Omega_k=-\Delta_k/2(k+1)$. It is not difficult to see that $\tau_k$ and $\Delta_k$ will increase synchronously with increasing $k$. Generally, the extension of gate time is undesirable for fast implementation of the operation. Nevertheless, in our scheme, the gate time can be controlled by adjusting the value of squeezing parameter $r_p$. That is to say, the reduction in gate time is expectable, which comes at a price of a corresponding increase in the parametric driving strength. By choosing separate squeezing parameters for the shaped gates of different orders, the corresponding gate times can be adjusted to be very close. Then, increasing $k$ can only result in the increase in $\Delta_k$, according to $\Delta_k=2(k+1)\pi/\tau_k$. Note that this increased detuning can suppress effectively the excitation of photon during the gate evolution. This is verified in Fig. \ref{fig3}(a) where we plot the time evolution of the mean photon number $\langle a^\dag a\rangle$ for both the unshaped gate and the shaped gates with $k$=1, 4 and 19. Apparently, a significant reduction in $\langle a^\dag a\rangle$ can be observed with increasing $k$. What is more intriguing is that the shaped gates display flatter evolutions of $\langle a^\dag a\rangle$ around the gate time in comparison with the unshaped gate. This stable and near-zero evolution of $\langle a^\dag a\rangle$ existing only in the shaped gates will decrease directly residual qutrit-cavity coupling that is caused by, e.g., gate timing error. 

A more intuitive understanding may be gained by examining the effect of using amplitude-shaped gate on the phase space trajectory. The propagator $U(t)$ leads to time-varying qubit-state-dependent  displacement in the $xp$ phase space of the cavity mode, and ideally (without errors), the trajectory closes at the gate time where the cavity and the qubits are disentangled, as shown in Fig. \ref{fig3}(b). Unlike the trajectory for the unshaped gate, with increasing $k$ the shaped-gates' trajectories display more windings with a reduced radius and become more centered around the origin. This effectively shortens the distance between the origin and the point in the phase space where the gate terminates in case of errors, thus mitigating the impact of errors due to residual entanglement between the cavity and the qubits.

\subsection{Control error in gate parameters}
\begin{figure}[t]
	\includegraphics[width=0.9\columnwidth]{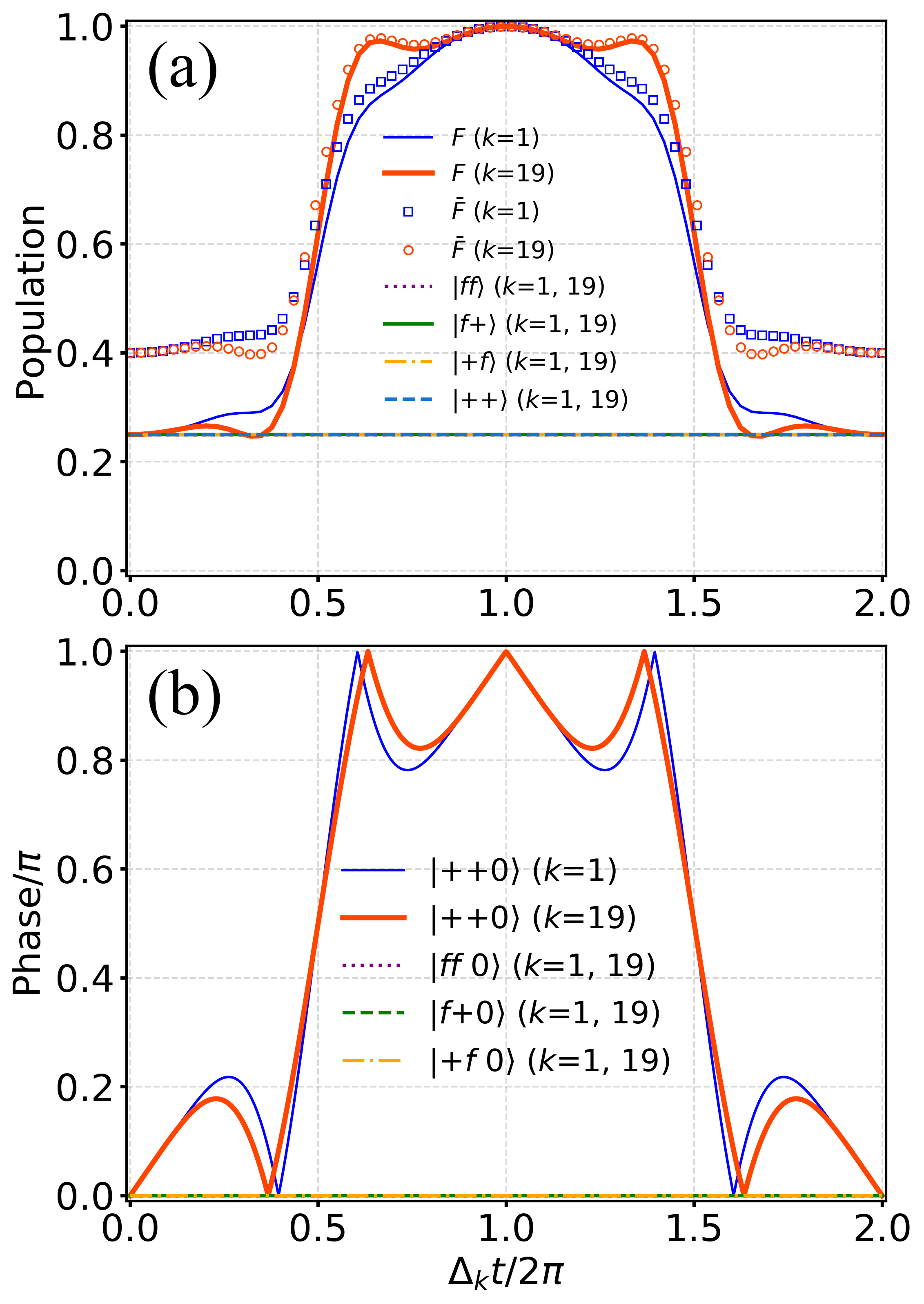}
	\caption{Population and fidelity evolutions (a) and phase evolutions (b) for the shaped gates with $k=1$ and 19. In (a), the initial-state-specified (average) gate fidelity for $k=1$ is marked with finer blue curve (hollow squares), while the initial-state-specified (average) gate fidelity for $k=19$ is marked with thicker red curve (hollow circles); The four straight lines representing population evolutions of different states overlap entirely. The common parameters used here are the same as in Fig .\ref{fig3}.}
	\label{fig4}
\end{figure}

To begin with, we briefly verify the validity of the proposed shaped gates. Taking the shaped gates with $k$=1 and 19 as examples, we plot in Fig. \ref{fig4} the corresponding population and phase evolutions of the computational states, as well as the evolutions of initial-state-specified and average fidelities. The introduction of the $\sin^2$ amplitude modulation results in a flatter response of fidelities around the gate time $\tau_k$ compared to the unshaped gate (see Fig. \ref{fig2}). Also, increasing $k$ will flatten the fidelity within a wider range, accompanied by smaller phase fluctuations around the gate time.

\begin{figure}
	\includegraphics[width=1\columnwidth]{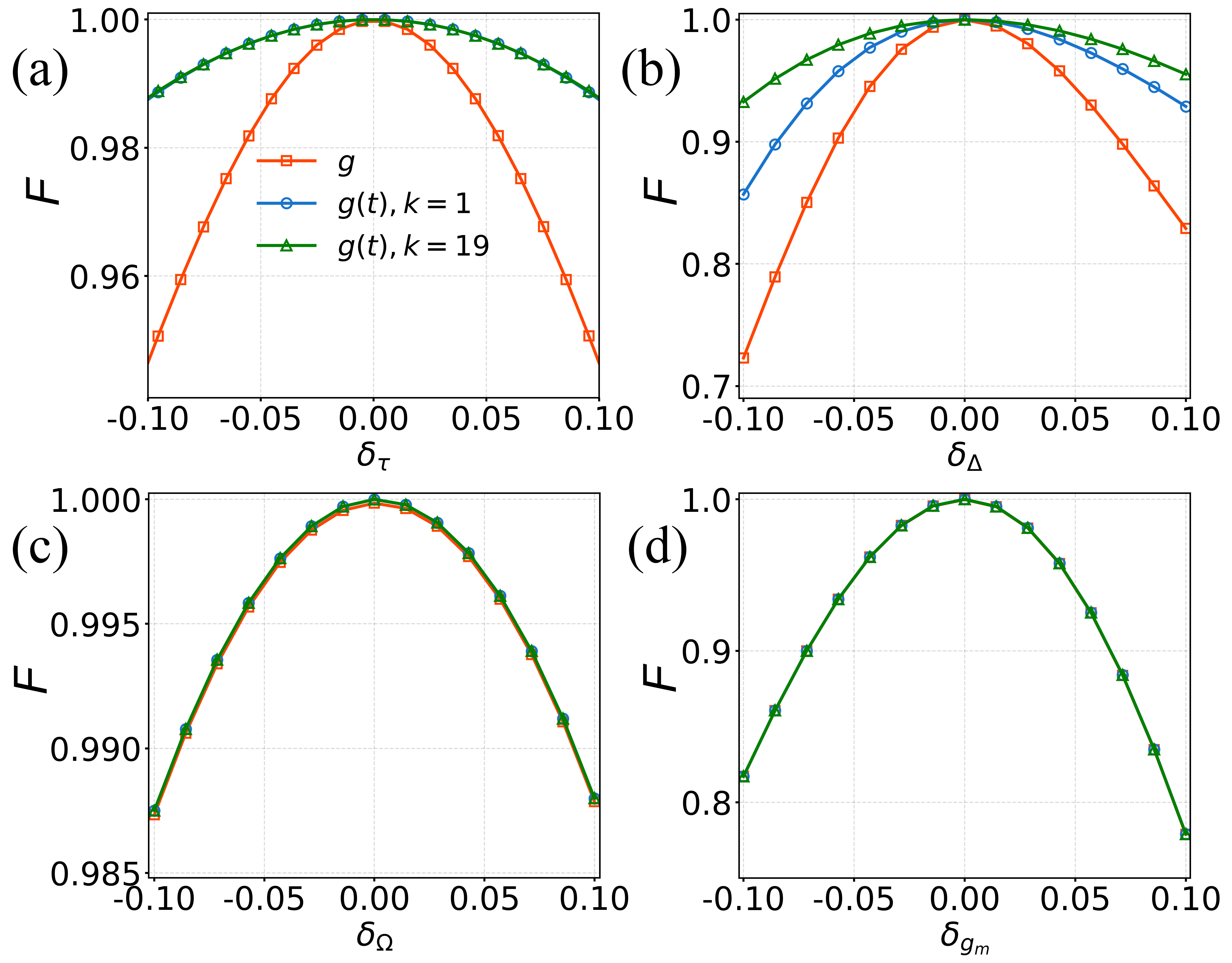}
	\caption{Gate fidelities versus control errors in (a) gate time $\tau$, (b) detuning $\Delta$, (c) classical-field amplitude $\Omega$, and (d) maximum amplitude of qutrit-cavity coupling strength $g_m$. The common parameters used here are the same as in Fig. \ref{fig3}.
	}
	\label{fig5}
\end{figure}

The above results show that stronger robustness against gate timing error is achievable. To be more clear, in Fig. \ref{fig5} we plot the gate fidelity as a function of four potential parameter errors for the unshaped gate and the shaped gates with $k=1$ and 19. The relative error of the parameter $x$ is defined as $\delta_x=(x'-x)/x$ in which $x'$ denotes the actual value and $x$ the ideal value. As seen in Fig. \ref{fig5}(a), the shaped gates display flatter response to $\delta_\tau$, indicating significant improvement of the gate robustness to the gate timing error. Increasing $k$ can hardly improve the gate robustness against $\delta_\tau$ within the given error range, but improve prominently the gate robustness against detuning error $\delta_\Delta$, as depicted in Fig. \ref{fig5}(b). For the amplitude error of the classical field $\delta_\Omega$, all three curves corresponding to different gates almost overlap entirely, and the identical fidelities keep always over 0.987 for $\delta_\Omega\in[-0.1,0.1]$. This manifests that slight deviation in $\Omega$ has small influence on our scheme. When considering the error in the maximum amplitude of the shaped coupling strength $\delta_{g_m}$, we find that even the fidelities of the shaped gates are also susceptible to $\delta_{g_m}$. High fidelity with $F>0.99$ is achievable provided that $\delta_{g_m}$ can be restrained in [-0.02,0.02].

\subsection{Effects of decoherence induced by dissipation}
We now take the effects of decoherence on the gate performance into account. Considering two dominant channels of decoherence induced by the dissipation of the system: (i) Energy relaxations of excited state of the qutrits; (ii) Decay of the cavity mode. Note that as we introduce a parametric driving to squeeze the cavity mode for achieving the Rabi model, the squeezing also inputs thermal noise and two-photon correlation noise into the cavity \cite{PhysRevLett.120.093601,PhysRevA.100.012339,PhysRevA.100.062501,PhysRevA.99.023833,doi:10.1002/andp.201900220,PhysRevLett.114.093602}. This results in dissipative dynamics that have a similar form to thermal dissipation. To eliminate completely these noises, we assume that the cavity is driven by a squeezed vacuum field, which can be implemented experimentally in circuit QED system using e.g., a Josephson parametric amplifier \cite{murch2013reduction,PhysRevLett.119.023602,doi:10.1063/1.2964182}. In doing so, the dynamics of the system could be described by the standard master equation
\begin{equation}\label{Eq:master equation} 	
\begin{split}
\dot{\rho}=i[\rho,H_s]+\sum_{z,k}\mathcal{L}(L_{zk})\rho+\mathcal{L}(L_{a})\rho,
\end{split}
\end{equation}
where $\rho$ is the density operator of the system, $H_s$ the full Hamiltonian given by Eq. (\ref{Hs}), $L_{zk}=\sqrt{\gamma}\vert z\rangle_k \langle e\vert$ ($z=g,f$) the Lindblad operators describing the energy relaxations of excited states with identical rate $\gamma$, $L_a=\sqrt{\kappa}a$ the Lindblad operator describing the cavity decay with rate $\kappa$, and $\mathcal{L}(o)\rho=o\rho o^\dag-(o^\dag o\rho+\rho o^\dag o)/2$. 

\begin{figure}
	\includegraphics[width=0.9\columnwidth]{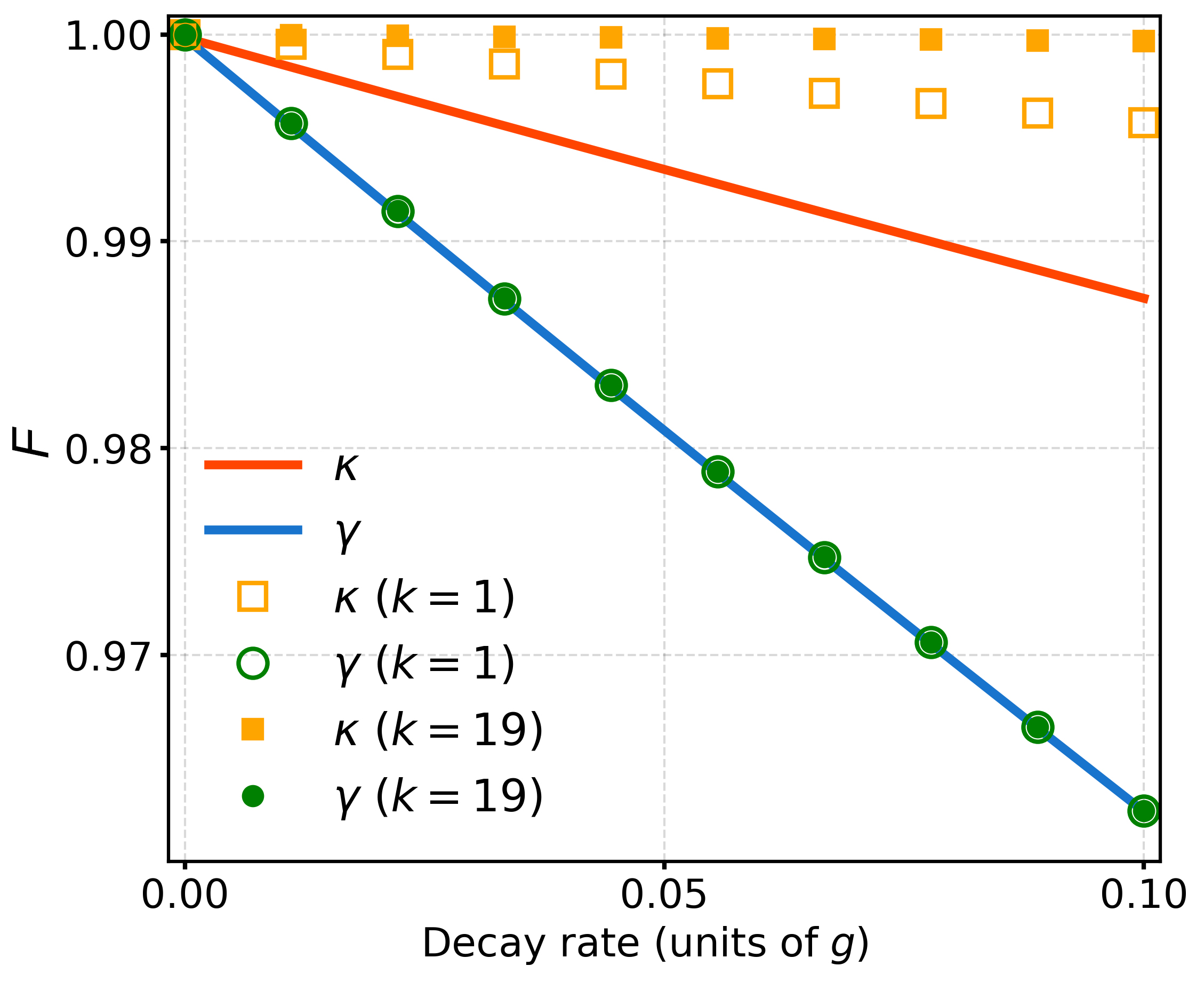}
	\caption{ Effects of cavity and qutrits decays on the gate fidelity for the unshaped gate (straight lines) and shaped gate with $k=1$ (hollow markers) and $k=19$ (solid markers). The common parameters used here are the same as in Fig. \ref{fig3}.
	}
	\label{fig6}
\end{figure}

Figure \ref{fig6} shows the gate fidelity (at the gate time) as a function of cavity or qutrit decay rate. It is shown that the three lines for $\gamma$ corresponding respectively to the unshaped and shaped gates almost overlap and the fidelities can still keep over 0.963 even for a large $\gamma$ of $0.1g$. The good robustness against qutrits decay is a consequence of coding quantum information on the qutrits ground states. In contrast, we find that the cavity decay influences less the gate fidelities. Particularly, for the shaped gate with a larger $k$ the gate fidelity is hardly reduced with increasing $\kappa$: the fidelity of the shaped gate with $k=19$ keeps over 0.9997 for $\kappa\le0.1g$. This remarkable enhancement of the gate robustness against cavity decay is because less excitation of the cavity mode occurs for the shaped gates with larger $k$.

\section{Discussion and conclusion}\label{sec4}
As discussed above, a most promising pathway for implementating our proposal could be the current circuit QED platform. The proposed amplitude-shaped gates are realized based on a unitary $U(t)$ with a specific evolution period. This requires the control of the time-dependent qutrit-resonator coupling $g(t)$, the classical field drives on the qutrits $\Omega$, and the parametric drive of the resonator $\Omega_p$, when considering the practical implementation in a superconducting circuit architecture. Specifically, the classical field drives could be implemented by introducing square microwave pulses with sine ramp-up and ramp-down edges and specific pulse length \cite{PhysRevLett.121.110501,PhysRevLett.124.230503}, while the parametric drive could be implemented through modulating the magnetic flux threading the SQUID loop connected to the resonator \cite{doi:10.1063/1.2964182,Zhong_2013}. Regarding the time-dependent coupling $g(t)$, there have been both theoretical \cite{PhysRevA.91.063814,PhysRevA.93.063845} and experimental \cite{PhysRevLett.104.177004,PhysRevLett.106.083601} research concerning coherent tunable qubit-resonator coupling. Experimentally, the desired amplitude modulation on $g(t)$ could be implemented directly by adopting controlled voltage pulses generated by an arbitrary waveform generator (AWG) to tune the flux threading the SQUID loop of each qutrits \cite{PhysRevX.5.021027}. All drives on qutrits could be set to respective pulse envelopes and be performed synchronously for a specific duration.

We next give a brief evaluation of the gate performance using experimentally accessible parameters. For a flux qubit coupled to a CPW resonator, it is demonstrated that the coupling strength up to several hundreds of MHz is achievable \cite{niemczyk2010circuit}. As an example, we assume a coupling strength of $g/2\pi=50$ MHz, which gives approximately $\Delta_1/2\pi =  556.05$ MHz and $\Omega_1/2\pi = -139.01$ MHz for our 1st order ($k=1$) shaped phase gate when considering a squeezing parameter of $r_p=2.5$, and the corresponding gate time is calculated as $\tau_1= 3.6$ ns. In addition, for the parameters assumed above, the other system parameters could be chosen as $\omega_a/2\pi = 43.76$ GHz, $\omega_p/2\pi = 5$ GHz, $(\omega_e-\omega_g)/2\pi = 2.5$ GHz, and $\omega/2\pi = 2.5$ GHz. Based on the above parameters, together with an assumption of experimentally feasible decay rates $\kappa/2\pi=\gamma/2\pi=0.5$ MHz \cite{yan2016flux,PhysRevLett.104.100504}, the shaped gate fidelity could reach above 0.99 at the gate time of about 3.6 ns.

\begin{figure}
	\includegraphics[width=0.9\columnwidth]{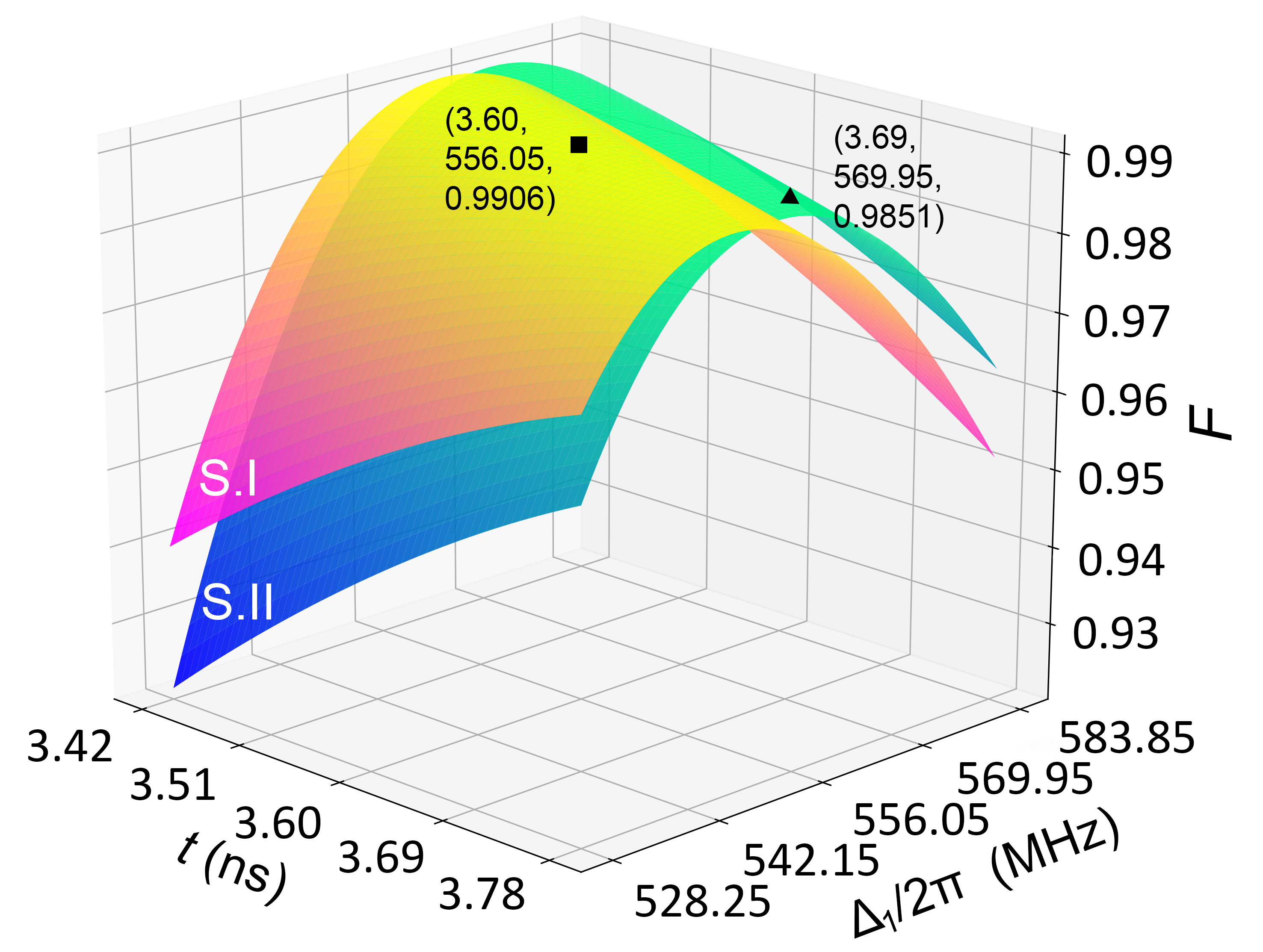}
	\caption{The 1st order shaped gate fidelity versus time $t$ and detuning $\Delta_1$. The parameters are: for the upper surface (S.I), $g_m/2\pi=50$ MHz, $\Omega_1/2\pi=-139.01$ MHz; for the lower surface (S.II), $g_m/2\pi=50.5$ MHz, $\Omega_1/2\pi=-140.4$ MHz. Other common parameters are: $\kappa/2\pi=\gamma/2\pi=0.5$ MHz.
	}
	\label{fig7}
\end{figure}

Regarding the effects of the potential errors in real experiments on the gate, we simulate the 1st order shaped gate fidelity versus varying actual values of $t$ and $\Delta_1$, as plotted in Fig. \ref{fig7}. The upper surface (labelled as ``S.I") corresponds to the case with perfect control of the pulse amplitude and the classical field amplitude, i.e., $\{g_m,\Omega_1\}$ precisely equals to $2\pi\times\{50,-139.01\}$ MHz, while the lower surface (labelled as ``S.II") depicts the case with slight deviation in these two amplitudes. Apparently, in the absence of all control errors, the shaped gate fidelity could achieve 0.9906, as marked by the square dot on S.I. When different control errors exist simultaneously, e.g., $t=3.69$ ns and $\{\Delta_1,g_m,\Omega_1\}=2\pi\times\{569.95,50.5,-140.4\}$ MHz, the corresponding gate fidelity would be reduced to 0.9851.

In conclusion, we have proposed a theoretical method for implementing fast two-qubit arbitrary phase gates and introduced amplitude modulation to improve the gate robustness against control errors and decoherence. The shaped gates provide enhanced robustness to errors in gate time and frequency detuning compared to the standard gate without amplitude-shaped coupling. In addition, the scheme presents intrinsic robustness against energy relaxations of the qutrits and in particular, significantly enhanced robustness against cavity decay is achieved with the shaped gates. Further improvement of the gate robustness would be possible by designing suitable pulse shapes with modulated amplitude and phase. We anticipate that our proposal could be helpful for future studies on this field and offer a feasible method for implementing efficient gate operations in quantum information protocols.

\section{Acknowledgements}
This work was supported by National Natural Science Foundation of China (NSFC) (11675046), Program for Innovation Research of Science in Harbin Institute of Technology (A201412), and Postdoctoral Scientific Research Developmental Fund of Heilongjiang Province (LBH-Q15060).

\section*{Appendix A: Effectiveness of the ideal Rabi Hamiltonian}
\begin{figure}
	\includegraphics[width=0.9\columnwidth]{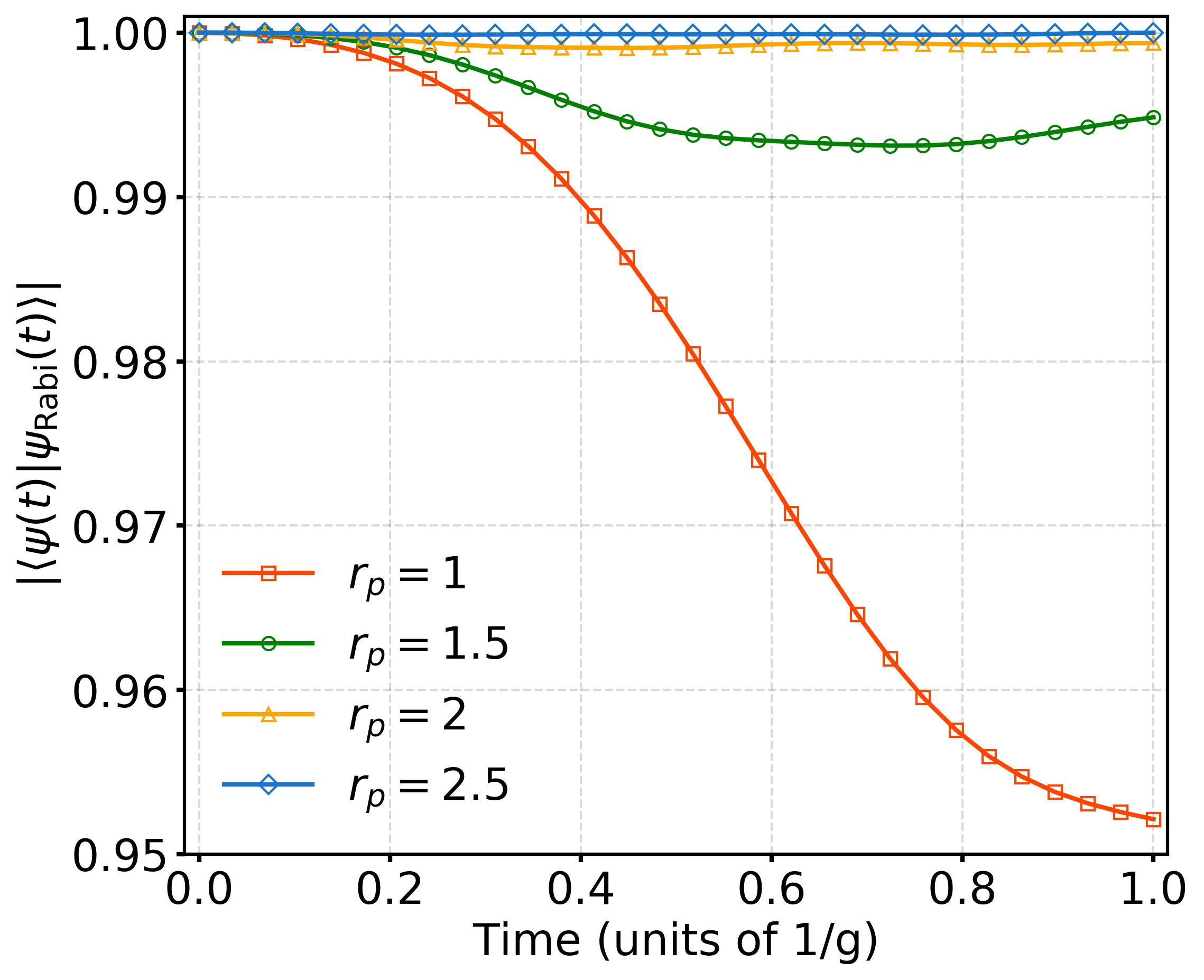}
	\caption{Evolution of the fidelity between the exact state $\psi(t)$ and the ideal Rabi-model state $\psi_{\mathrm{Rabi}}(t)$, which are obtained by solving the exact Hamiltonian ($H_s$) and the ideal Rabi Hamiltonian ($H_s$ in the absence of $H_{\mathrm{Err}}$). The parameters are: $\Delta=ge^{r_p}$ and $\Omega=-\Delta/2$.
	}
	\label{fig8}
\end{figure}

As seen in Eq. (\ref{Hs}), the error term $H_{\mathrm{Err}}= \frac{g}{2}e^{-r_p}(a^\dag-a)(\lvert e\rangle_k\langle g\lvert-\lvert g\rangle_k\langle e\lvert)$ describes undesired corrections to the ideal Rabi Hamiltonian. In principle, $H_{\mathrm{Err}}$ can be neglected completely for a large $r_p$ satisfying $e^{r_p}\to\infty$. Here we demonstrate numerically that a modest $r_p$ is able to supress $H_{\mathrm{Err}}$ sufficiently. To be specific, we plot in Fig. \ref{fig8} the dynamical evolution of the fidelity between the exact state $\psi(t)$ and the ideal Rabi-model state $\psi_{\mathrm{Rabi}}(t)$ for several values of $r_p$, where $\psi(t)$ and $\psi_{\mathrm{Rabi}}(t)$ are obtained by solving the Hamiltonian (\ref{Hs}) with and without $H_{\mathrm{Err}}$, respectively. Apparently, the fidelity keeps near unity over long time scales for $r_p=2.5$, which implies that the system described by the exact Hamiltonian faithfully reproduces the ideal Rabi-model evolution. This allows us to choose $r_p=2.5$ in the simulations, as shown in Fig. \ref{fig2}, where a high-fidelity gate can be achieved correspondingly. 

\bibliography{sample.bib}

\end{document}